# Non-Hermitian Linear Electrooptic Effect in 3D materials

Tiago A. Morgado[1], Tatiana G. Rappoport[2,3], Stepan S. Tsirkin[4,5], Sylvain Lannebère[1], Ivo Souza[4,5], and Mário G. Silveirinha[6*]

[1]*Instituto de Telecomunicações and Department of Electrical Engineering, University of Coimbra, 3030-290 Coimbra, Portugal*

[2]*Centro de Física das Universidades do Minho e do Porto (CF-UM-UP) e Departamento de Física, Universidade do Minho, P-4710-057 Braga, Portugal*

[3]*Instituto de Física, Universidade Federal do Rio de Janeiro, C.P. 68528, 21941-972 Rio de Janeiro RJ, Brazil*

[4]*Centro de Física de Materiales, Universidad del País Vasco, 20018 San Sebastián, Spain*

[5]*Ikerbasque Foundation, 48013 Bilbao, Spain*

[6]*University of Lisbon–Instituto Superior Técnico and Instituto de Telecomunicações, Avenida Rovisco Pais, 1, 1049-001 Lisboa, Portugal*

*E-mail:* tiago.morgado@co.it.pt, mario.silveirinha@tecnico.ulisboa.pt

**Abstract**

Here, we present an in-depth theoretical analysis of the linear electrooptic effect in low-symmetry three-dimensional (3D) conductive materials with large Berry curvature dipoles. Our study identifies two distinct kinetic contributions to the linear electrooptic effect: a gyrotropic Hermitian (conservative) piece and a non-Hermitian term that can originate optical gain. We concentrate on the study of 3D materials belonging to the 32 ($\mathcal{D}_3$) point group subject to a static electric bias along the trigonal axis. Our investigation shows that doped trigonal tellurium has promising properties, with its gyrotropic electrooptic response offering the potential for realizing electrically-biased electromagnetic isolators and inducing significant optical dichroism. Most notably, it is demonstrated that under sufficiently large static electric bias, tellurium's non-Hermitian electrooptic response may lead to optical gain. Using first-principles calculations, it is shown that *n*-doped tellurium is particularly promising, as it can host significantly larger Berry curvature dipoles than the more common *p*-doped tellurium.

---

[*] To whom correspondence should be addressed: E-mail: tiago.morgado@co.it.pt

# I. INTRODUCTION

Advances in integrated photonic circuits are key to meeting the demands of high-speed data communications, offering significant improvements in speed, bandwidth, and energy efficiency [1-2]. However, their development is challenged by the presence of losses and the difficulty of integrating nonreciprocal components [1, 3-7]. Nonreciprocal components, such as isolators and circulators, enable one-way light transmission, a crucial feature for efficient signal routing and control in photonic circuits.

The traditional and long-established approach to achieve robust nonreciprocal responses and realize electromagnetic isolators involves breaking time-reversal symmetry by exploiting magneto-optic effects in materials like ferrites or iron garnets under a static magnetic bias [8-10]. However, the requirement for an external bulky magnetic biasing circuit poses a major obstacle to integrating these components onto a chip. This issue has spurred the development of alternative "magnetless" nonreciprocal systems in recent years.

Magnetless nonreciprocal systems can be categorized into two distinct groups. The first group comprises systems with linear material responses under normal operating conditions, and which require the application of a suitable external bias. This group includes various platforms characterized by broken time-reversal symmetry, such as time-variant systems [11-14], systems with moving parts [15-16], and systems with drifting electrons [17-22]. Furthermore, this category also encompasses proposals involving non-Hermitian platforms based on active electronic systems [23-25] or optically pumped systems [26]. The second group comprises nonreciprocal systems that rely on the use of nonlinear materials, which are dynamically self-biased by incoming waves [27-32]. The latter solutions require high-power input signals and usually do not offer robust optical isolation [33-34].

Recently [35], inspired by the physics of transistors, we introduced a novel mechanism for generating strongly nonreciprocal and non-Hermitian linearized electromagnetic responses in low-symmetry materials. Our approach relies on the combination of material nonlinearities with a static electric bias. In our theoretical work [35], we demonstrated that a hypothetical metamaterial, composed of a periodic arrangement of MOSFETs (termed "MOSFET-metamaterial") exhibits intriguing nonreciprocal and non-Hermitian responses with unique physical properties under a static electric bias.

Our theory raises the question of whether it is possible to achieve a distributed electromagnetic response akin to that of a transistor using natural materials. In a recent theoretical study [36], we have demonstrated that nonlinear two-dimensional (2D) conductive materials with a broken inversion symmetry and a large Berry curvature dipole (e.g., strained twisted bilayer graphene) are promising platforms to obtain nonreciprocal and non-Hermitian transistor-like distributed responses. The Berry curvature dipole may be pictured as a dipolar pattern of the Berry curvature (distribution) across the Fermi surface. Our analysis shows that an incident electromagnetic wave passing through such electrically-biased 2D materials may experience optical gain, depending on the wave polarization, direction of wave propagation, and orientation of the static electric bias.

In this article, we extend our theory to three-dimensional (3D) crystals with broken inversion symmetry and a Berry curvature dipole. Consistent with our previous investigation on 2D low-symmetry materials [36], we find that there are two distinct contributions to the linear electrooptic effect in these 3D materials: one is rooted in a gyrotropic Hermitian (conservative) response, and the other is associated with a non-Hermitian electrooptic response characterized by its non-conservative nature. Moreover, we analyze the origin for the gyrotropic electrooptic effect, suggesting that it arises from microscopic DC currents following helical-type trajectories induced by the static electric bias, which in turn generate an internal static magnetic field.

Throughout this work, we focus the study of the linear electrooptic effects on 3D materials belonging to the 32-point group ($\mathcal{D}_3$ in the Schoenflies notation) subject to a static electric bias along the trigonal axis. Using first principles density functional theory (DFT) calculations, we provide an in-depth study of the linear electrooptic effect in trigonal tellurium. It is demonstrated that the gyrotropic Hermitian response in *n*-doped tellurium may present exciting opportunities, e.g., the realization of electrically-biased electromagnetic isolators and the generation of significant optical dichroism. Indeed, the first principles calculations suggest that *n*-doped tellurium may provide larger Berry curvature dipoles and lower DC conductivity in comparison with the more common *p*-doped variant. Moreover, under a sufficiently high static electric bias, the non-Hermitian electrooptic response of tellurium may give rise to optical gain.

This article is organized as follows. In Sec. II, we derive the gyrotropic (Hermitian) and non-Hermitian linear electrooptic (EO) responses of generic 3D low-symmetry conductive systems. In Sec. III, we examine the linear EO effect in generic 3D materials belonging to the 32-point group symmetry ($\mathcal{D}_3$). Finally, in Sec. IV we investigate nonreciprocal and gain effects in trigonal tellurium biased along the trigonal axis. The conclusions are drawn in Sec. V.

## II. NON-HERMITIAN LINEAR ELECTROOPTIC EFFECT IN LOW-SYMMETRY MATERIALS

### A. Boltzmann transport theory

Here, we use the semiclassical Boltzmann transport theory to derive the linearized optical response of a generic low-symmetry metallic system under a static electric bias. We suppose that the material is described by a time-reversal symmetric Hamiltonian.

The electron distribution function in the material, $f_\mathbf{k}$, satisfies the Boltzmann transport equation:

$$\frac{\partial f_\mathbf{k}}{\partial t} + \frac{\partial \mathbf{k}}{\partial t} \cdot \nabla_\mathbf{k} f_\mathbf{k} = -\frac{1}{\tau}\left(f_\mathbf{k} - f_\mathbf{k}^0\right), \tag{1}$$

where $f_\mathbf{k}^0$ is the Fermi-Dirac distribution, $\tau$ is the scattering relaxation time, and $\hbar \frac{\partial \mathbf{k}}{\partial t} = -e\mathbf{E}$ (with $e > 0$). For a static (DC) field $\mathbf{E}_0$ the distribution function does not depend on $t$. Thus, the perturbation due to the static bias is given by:

$$\left[f_\mathbf{k} - f_\mathbf{k}^0\right]_{\text{static}} = \delta f_\mathbf{k}^0 \approx \frac{e}{\hbar}\tau \mathbf{E}_0 \cdot \nabla_\mathbf{k} f_\mathbf{k}^0. \tag{2}$$

Let us now suppose that a much weaker optical dynamic (AC) field ($\mathbf{E}_\omega e^{-i\omega t}$) is applied to the material, resulting in a perturbation ($\delta f_\mathbf{k}^\omega e^{-i\omega t}$) of the distribution function. The linearized dynamics of the optical field is controlled by the distribution function $f_\mathbf{k}^0 + \delta f_\mathbf{k}^0$, so that $\delta f_\mathbf{k}^\omega$ satisfies:

$$\frac{\partial}{\partial t}\left(\delta f_\mathbf{k}^\omega e^{-i\omega t}\right) + \left(-e\mathbf{E}_\omega e^{-i\omega t}\right)\frac{1}{\hbar}\cdot\nabla_\mathbf{k}\left[f_\mathbf{k}^0 + \delta f_\mathbf{k}^0\right] = -\frac{1}{\tau}\delta f_\mathbf{k}^\omega e^{-i\omega t}, \tag{3}$$

Solving Eq. (3) with respect to $\delta f_\mathbf{k}^\omega$ it follows that:

$$\begin{aligned}\delta f_\mathbf{k}^\omega &= \frac{1}{1-i\omega\tau}\frac{\tau e}{\hbar}\nabla_\mathbf{k}\left[f_\mathbf{k}^0 + \delta f_\mathbf{k}^0\right]\cdot\mathbf{E}_\omega \\ &= \frac{1}{1-i\omega\tau}\frac{\tau e}{\hbar}\nabla_\mathbf{k} f_\mathbf{k}^0 \cdot\mathbf{E}_\omega + \frac{1}{1-i\omega\tau}\left(\frac{\tau e}{\hbar}\right)^2 \nabla_\mathbf{k}\left[\mathbf{E}_0\cdot\nabla_\mathbf{k} f_\mathbf{k}^0\right]\cdot\mathbf{E}^\omega\end{aligned} \tag{4}$$

Next, we characterize the induced current density distribution due to the applied fields. It can be written in terms of the total distribution function as follows:

$$\mathbf{J} = -\frac{e}{V}\sum_{\mathbf{k}}\left(f_{\mathbf{k}}^0 + \delta f_{\mathbf{k}}^0 + \delta f_{\mathbf{k}}^\omega e^{-i\omega t} + c.c.\right)\left(\mathbf{v}_{\mathbf{k}}^0 - \frac{e}{\hbar}\mathbf{\Omega}_{\mathbf{k}}\times(\mathbf{E}_0 + \mathbf{E}_\omega e^{-i\omega t} + c.c.)\right), \quad (5)$$

where $\mathbf{v}_{\mathbf{k}}^0$ is the group velocity of the electron wave, $\mathbf{\Omega}_{\mathbf{k}}$ is the Berry curvature, $V$ is the volume of the material, and *c.c.* stands for complex conjugate. Note that we take into account the anomalous velocity contribution, $-\frac{e}{\hbar}\mathbf{\Omega}_{\mathbf{k}}\times\mathbf{E}$ [37-38], which is the key mechanism to obtain both the non-Hermitian and the gyrotropic response. The anomalous velocity has higher-order corrections in the electromagnetic fields [39], but they have no impact in the EO response of time-reversal symmetric systems. The linear response is determined by the following static and dynamic current densities

$$\mathbf{J}^0 = -\frac{e}{V}\sum_{\mathbf{k}}\left(\mathbf{v}_{\mathbf{k}}^0 - \frac{e}{\hbar}\mathbf{\Omega}_{\mathbf{k}}\times\mathbf{E}_0\right)\left(f_{\mathbf{k}}^0 + \delta f_{\mathbf{k}}^0\right), \quad (6a)$$

$$\mathbf{J}^\omega = -\frac{e}{V}\sum_{\mathbf{k}}\left(\mathbf{v}_{\mathbf{k}}^0 - \frac{e}{\hbar}\mathbf{\Omega}_{\mathbf{k}}\times\mathbf{E}_0\right)\delta f_{\mathbf{k}}^\omega - \frac{e}{V}\sum_{\mathbf{k}}\left(-\frac{e}{\hbar}\mathbf{\Omega}_{\mathbf{k}}\times\mathbf{E}_\omega\right)\left(f_{\mathbf{k}}^0 + \delta f_{\mathbf{k}}^0\right). \quad (6b)$$

As the system is described by a time-reversal symmetric Hamiltonian, $\mathbf{v}_{\mathbf{k}}^0$ and $\mathbf{\Omega}_{\mathbf{k}}$ are odd functions of the quasi-momentum $\mathbf{k}$. Thus, only the terms associated with $\delta f_{\mathbf{k}}^0$ and $\delta f_{\mathbf{k}}^\omega$, which are also odd functions of $\mathbf{k}$, can contribute to the response:

$$\mathbf{J}_0 = -\frac{e}{V}\sum_{\mathbf{k}}\left(\mathbf{v}_{\mathbf{k}}^0 - \frac{e}{\hbar}\mathbf{\Omega}_{\mathbf{k}}\times\mathbf{E}_0\right)\delta f_{\mathbf{k}}^0, \quad (7a)$$

$$\mathbf{J}_\omega = -\frac{e}{V}\sum_{\mathbf{k}}\left(\mathbf{v}_{\mathbf{k}}^0 - \frac{e}{\hbar}\mathbf{\Omega}_{\mathbf{k}}\times\mathbf{E}_0\right)\delta f_{\mathbf{k}}^\omega - \frac{e}{V}\sum_{\mathbf{k}}\left(-\frac{e}{\hbar}\mathbf{\Omega}_{\mathbf{k}}\times\mathbf{E}_\omega\right)\delta f_{\mathbf{k}}^0. \quad (7b)$$

Using now Eqs. (2) and (4), we can write the current as a function of the applied electric field:

$$\mathbf{J}_0 = -\frac{e}{V}\sum_{\mathbf{k}}\mathbf{v}_{\mathbf{k}}^0\left(\frac{\tau e}{\hbar}\mathbf{E}_0\cdot\nabla_{\mathbf{k}}f_{\mathbf{k}}^0\right) - \frac{e}{V}\sum_{\mathbf{k}}\left(-\frac{e}{\hbar}\mathbf{\Omega}_{\mathbf{k}}\times\mathbf{E}_0\right)\left(\frac{\tau e}{\hbar}\mathbf{E}_0\cdot\nabla_{\mathbf{k}}f_{\mathbf{k}}^0\right), \quad (8a)$$

$$\mathbf{J}_\omega = -\frac{e}{V}\sum_{\mathbf{k}} \mathbf{v}_{\mathbf{k}}^0 \left( \frac{1}{1-i\omega\tau} \frac{\tau e}{\hbar} \nabla_{\mathbf{k}} f_{\mathbf{k}}^0 \cdot \mathbf{E}_\omega \right) - \frac{e}{V}\sum_{\mathbf{k}} \left( -\frac{e}{\hbar} \mathbf{\Omega}_{\mathbf{k}} \times \mathbf{E}_0 \right) \left( \frac{1}{1-i\omega\tau} \frac{\tau e}{\hbar} \nabla_{\mathbf{k}} f_{\mathbf{k}}^0 \cdot \mathbf{E}_\omega \right)$$
$$-\frac{e}{V}\sum_{\mathbf{k}} \left( \frac{\tau e}{\hbar} \mathbf{E}_0 \cdot \nabla_{\mathbf{k}} f_{\mathbf{k}}^0 \right) \left( -\frac{e}{\hbar} \mathbf{\Omega}_{\mathbf{k}} \times \mathbf{E}_\omega \right)$$

(8b)

We only retain terms of $\mathbf{J}_\omega$ that are either independent of $\mathbf{E}_0$ or linear in $\mathbf{E}_0$. Higher-order terms are weaker and are not accounted for in our analysis. The leading terms in (8) associated with $\mathbf{v}_{\mathbf{k}}^0$ are the usual linear static and dynamic Drude-like responses. The linear EO response is ruled by the remaining terms:

$$\mathbf{J}_{EO} = \frac{\tau e^3}{\hbar^2} \mathbf{E}^0 \cdot \left( \frac{1}{V} \sum_{\mathbf{k}} \nabla_{\mathbf{k}} f_{\mathbf{k}}^0 \otimes \mathbf{\Omega}_{\mathbf{k}} \right) \times \mathbf{E}_\omega$$
$$\frac{\tau e^3}{\hbar^2} \frac{1}{1-i\omega\tau} \left( \frac{1}{V} \sum_{\mathbf{k}} (\mathbf{\Omega}_{\mathbf{k}} \times \mathbf{E}_0) \otimes \nabla_{\mathbf{k}} f_{\mathbf{k}}^0 \right) \cdot \mathbf{E}_\omega$$

(9)

where the symbol $\otimes$ denotes the tensor product of two vectors. The EO current density ($\mathbf{J}_{EO}$) is ruled by the anomalous electron velocity. Specifically, it arises from the linear variation of the distribution function under the influence of one of the applied fields ($\mathbf{E}_0$ or $\mathbf{E}_\omega$), in conjunction with the anomalous velocity determined by the interplay between the Berry curvature and the other applied field ($\mathbf{E}_\omega$ or $\mathbf{E}_0$).

### B. Conductivity response

The linear electrooptic response is determined by Eq. (9). The corresponding conductivity tensor can be written as a Brillouin-zone integral using the rule $\frac{1}{V}\sum_{\mathbf{k}} \to \frac{1}{(2\pi)^3} \int d^3\mathbf{k}$. Thus, the linear EO piece of the conductivity can be written as:

$$\bar{\sigma}_{EO}(\omega) = \frac{\tau e^3}{\hbar^2}\left(\mathbf{E}_0 \cdot \frac{1}{(2\pi)^3}\int d^3\mathbf{k}\, \nabla_\mathbf{k} f_\mathbf{k}^0 \otimes \mathbf{\Omega}_\mathbf{k}\right) \times \bar{\mathbf{1}}$$
$$-\frac{\tau e^3}{\hbar^2}\frac{1}{1-i\omega\tau}\mathbf{E}_0 \times \left(\frac{1}{(2\pi)^3}\int d^3\mathbf{k}\left(\mathbf{\Omega}_\mathbf{k} \otimes \nabla_\mathbf{k} f_\mathbf{k}^0\right)\right), \quad (10)$$

with $\bar{\mathbf{1}}$ the unit matrix. The conductivity can be expressed in terms of the so-called Berry curvature dipole (BD) tensor ($\bar{\mathbf{D}}$) defined as:

$$D_{ij} = -\frac{1}{(2\pi)^3}\int \frac{\partial f_\mathbf{k}^0}{\partial k_i}\Omega_{\mathbf{k},j}d^3\mathbf{k} = \frac{1}{(2\pi)^3}\int f_\mathbf{k}^0 \frac{\partial \Omega_{\mathbf{k},j}}{\partial k_i}d^3\mathbf{k}. \quad (11)$$

The Berry curvature dipole in 3D materials is dimensionless and traceless: $D_{xx} + D_{yy} + D_{zz} = 0$. The conductivity $\bar{\sigma}_{EO}$ is given by:

$$\bar{\sigma}_{EO}(\omega) = \bar{\sigma}_{EO}^{H} + \bar{\sigma}_{EO}^{NH}(\omega)$$
$$\equiv \frac{-\tau e^3}{\hbar^2}(\mathbf{E}_0 \cdot \bar{\mathbf{D}}) \times \bar{\mathbf{1}} + \frac{\tau e^3}{\hbar^2}\frac{1}{1-i\omega\tau}(\mathbf{E}_0 \times \bar{\mathbf{D}}^T), \quad (12)$$

where the symbol $T$ denotes the tensor transpose.

The first term of $\bar{\sigma}_{EO}$ defines an anti-symmetric real-valued matrix, analogous to a Hall conductivity. It corresponds to a conservative gyrotropic response associated with a pseudo-vector aligned along the direction of the vector $\mathbf{E}_0 \cdot \bar{\mathbf{D}} = \bar{\mathbf{D}}^T \cdot \mathbf{E}_0$ (pseudo-magnetic field). This gyrotropic conservative response, which has been previously discussed in the literature [40], plays a crucial role in multiple phenomena, including the kinetic Faraday effect (also known as current-induced optical activity) [40-43] and the kinetic magnetoelectric effect [40, 44-45]. Remarkably, $\bar{\sigma}_{EO}$ has an additional contribution ($\bar{\sigma}_{EO}^{NH}$) that varies with frequency and generally leads to a non-conservative electrooptic response and optical gain. This non-Hermitian electrooptic response was discovered only recently [36, 46]. It should be noted that the derived EO response [Eq. (12)] is only valid for frequencies well below the threshold for interband absorption.

## C. The Berry curvature dipole as a magneto-electric coupling tensor

To provide some intuition on the origin of the electrooptic effect and on the physical interpretation of the Berry curvature dipole from an optics perspective, next we present a comparison between the linear EO response of (chiral) conductors and the response of magnetized (gyrotropic) materials.

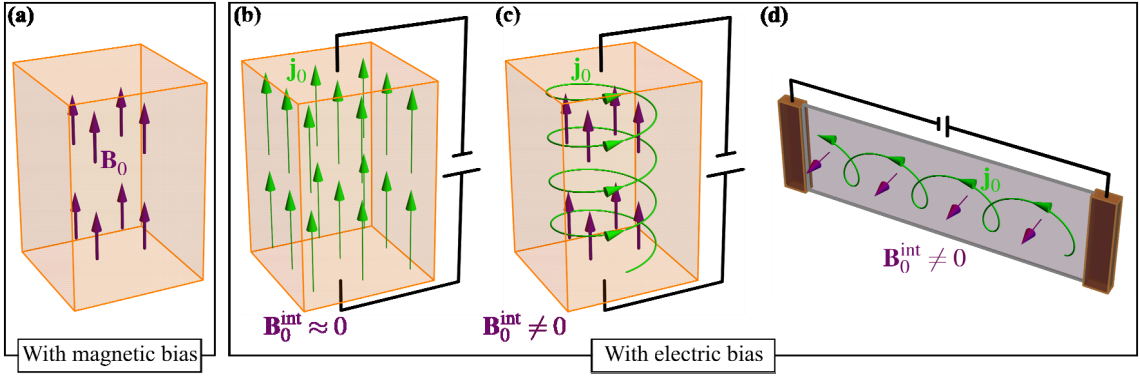

**Fig. 1**. (a) Electron gas biased by a static magnetic field $\mathbf{B}_0$. The bias induces a gyrotropic electromagnetic response. (b) Semiconductor with mirror symmetry biased with a static electric field: the average magnetic field in the bulk region is approximately zero, resulting in a negligible gyrotropic response. (c) Chiral low-symmetry semiconductor biased with a static electric field ($\mathbf{E}_0$). Due to the low-symmetry of the material, the electric current $\mathbf{j}_0$ may follow a helical-type path, which creates an internal magnetic field and a gyrotropic electromagnetic response. (d) Similar to (c) but for a 2D system. In this case, the current trajectory has the form of a planar-helix.

To this end, let us start by analyzing the response of an electron gas biased with a static magnetic field [Fig. 1(a)]. A magnetized electron gas exhibits a gyrotropic response, which can be modeled by a conductivity tensor of the form $\bar{\sigma}_{\text{gyro}} \approx \varepsilon_0 \frac{\omega_p^2}{\Gamma^2} \boldsymbol{\omega}_c \times \bar{\mathbf{1}}$, with $\varepsilon_0$ the vacuum permittivity, $\omega_p$ the plasma frequency, $\boldsymbol{\omega}_c = e\mathbf{B}_0/m^*$ the cyclotron frequency, and $\Gamma = 1/\tau$ the collision frequency associated with material loss. For simplicity, we focus on the low-frequency response of the material ($\omega \ll \Gamma$) and suppose that $|\omega_c| \ll \Gamma$.

It is interesting to compare the response of the magnetically-biased electron gas with the first piece of Eq. (12). By matching $\bar{\sigma}_{\text{gyro}} \approx \varepsilon_0 \frac{\omega_p^2}{\Gamma^2} \boldsymbol{\omega}_c \times \bar{\mathbf{1}}$ to $\bar{\sigma}_{\text{EO}}^{H} = -\frac{\tau e^3}{\hbar^2}(\mathbf{E}_0 \cdot \bar{\mathbf{D}}) \times \bar{\mathbf{1}}$ one sees that the static electric bias induces an equivalent cyclotron frequency given by $\boldsymbol{\omega}_c = \frac{-e^3 \Gamma}{\varepsilon_0 \hbar^2 \omega_p^2}(\bar{\mathbf{D}}^T \cdot \mathbf{E}_0)$. Thus, the equivalent magnetic bias is given by:

$$\mathbf{B}_0^{\text{int}} = \frac{-e^2 \Gamma m^*}{\varepsilon_0 \hbar^2 \omega_p^2} \bar{\mathbf{D}}^T \cdot \mathbf{E}_0 = -\frac{1}{c}\bar{\zeta} \cdot \mathbf{E}_0, \quad (13)$$

where $m^*$ is the effective mass of the electrons, and $\bar{\zeta} = 4\pi\alpha_e \frac{m^* c^2}{\hbar \omega_p} \frac{\Gamma}{\omega_p} \bar{\mathbf{D}}^T$ is a dimensionless tensor, with $\alpha_e = \frac{e^2}{4\pi\varepsilon_0 \hbar c} \approx \frac{1}{137}$ the fine-structure constant. As seen, the Berry curvature dipole links the equivalent magnetic field to the static electric bias. Thus, $\bar{\mathbf{D}}^T$ may be understood as a static magneto-electric coupling tensor. A related interpretation was presented in Ref. [47]. To give an idea of the numbers, suppose that $\Gamma/\omega_p \sim 1$, $m^* = 0.1 m_e$ with $m_e$ the rest free-electron mass, $\omega_p \sim 2\pi \times 1$ THz, and $\bar{\mathbf{D}}^T \sim 10^{-4}$. Then, the electric field intensity required in order that $\mathbf{B}_0^{\text{int}} \sim 1$ mT is $2.6 \times 10^3$ V/m. The equivalent magnetic field intensity scales with the Berry curvature dipole, and hence materials with a large Berry curvature dipole can be promising platforms to generate strong nonreciprocal and non-Hermitian effects.

The previous discussion leads to an interesting picture for the physical origin of the (conservative part) of the linear EO effect. In fact, it suggests that in low-symmetry materials the static electric bias may create an internal static magnetic field that acts to bias the material. The effect may be pictured as a result of helical-type trajectories for the DC current, due to the low-symmetry of the material [compare Figs. 1(b) and 1(c)]. In contrast, for materials with mirror-symmetry the current trajectory follows a straight line, and thereby the induced magnetic field is mostly outside of the material and the net magnetization within the material is approximately zero (Fig. 1(b)). Consequently, the gyrotropic response is negligible in the high-symmetry case ($\bar{\sigma}_{\text{gyro}} \approx 0$). A related picture for the current trajectory in the material has been discussed elsewhere [48-50]. The same mechanism justifies the gyrotropic response of low-symmetry 2D materials biased with a static electric bias (see Ref. [36]). In this case, the microscopic currents may flow along planar helical paths, leading to a magnetic field orthogonal to $\mathbf{E}_0$ [see Fig. 1(d)].

Importantly, the current-induced gyrotropic effect has been experimentally observed in Refs. [41-42], by measuring the rotation of light polarization when passing through a biased tellurium sample. This electrical analogue of the Faraday effect [51] was designated as current-induced optical activity [41-42] and kinetic Faraday effect [40].

## III. MATERIALS WITH $\mathcal{D}_3$ SYMMETRY

### A. Electrooptic conductivity

Here we analyze the electrooptic effects in 3D materials belonging to the 32 point group. For such materials, the Berry curvature dipole tensor has the form [40]:

$$\bar{\mathbf{D}} = D\hat{\mathbf{x}} \otimes \hat{\mathbf{x}} + D\hat{\mathbf{y}} \otimes \hat{\mathbf{y}} - 2D\hat{\mathbf{z}} \otimes \hat{\mathbf{z}}. \tag{14}$$

For a static electric field bias $\mathbf{E}_0 = E_0\hat{\mathbf{z}}$ applied along the trigonal axis, the Hermitian and non-Hermitian EO conductivity contributions are given by:

$$\bar{\sigma}_{\text{EO}}^{\text{H}} = \frac{\tau e^3}{\hbar^2} 2DE_0 \hat{\mathbf{z}} \times \bar{\mathbf{1}}, \tag{15a}$$

$$\bar{\sigma}_{\text{EO}}^{\text{NH}}(\omega) = \frac{\tau e^3}{\hbar^2} \frac{DE_0}{1-i\omega\tau} \hat{\mathbf{z}} \times \bar{\mathbf{1}}. \tag{15b}$$

Curiously, for a $\mathcal{D}_3$-symmetry, both conservative and non-conservative components exhibit identical structures: they are represented by an anti-symmetric tensor. As further discussed below, the response described by $\bar{\sigma}_{EO}^{NH}(\omega)$ is inherently non-conservative.

For convenience we introduce an equivalent cyclotron-type frequency defined as:

$$\omega_0 = \frac{\tau e^3}{\varepsilon_0 \hbar^2} DE_0 = 4\pi \alpha_e \frac{ec}{\hbar \Gamma} DE_0. \tag{16}$$

The frequency $\omega_0$ is proportional to the dipole strength and to the electric field strength, and inversely proportional to the collision frequency ($\Gamma = 1/\tau$). The linear electrooptic conductivity can be written in terms of $\omega_0$ as follows:

$$\bar{\sigma}_{EO}^{H} = 2\varepsilon_0 \omega_0 \hat{\mathbf{z}} \times \overline{\mathbf{1}}, \qquad \bar{\sigma}_{EO}^{NH}(\omega) = \frac{\varepsilon_0 \omega_0}{1 - i\omega\tau} \hat{\mathbf{z}} \times \overline{\mathbf{1}}. \tag{17}$$

It is interesting to analyze the EO contribution to the power (per unit of volume) transferred from the optical field to the material (dissipated power). It is given by:

$$p_{dis,EO} = \frac{1}{2} \text{Re}\{\mathbf{E}^* \cdot \bar{\sigma}_{EO} \cdot \mathbf{E}\} = -\varepsilon_0 \omega_0 \frac{\omega \Gamma}{\omega^2 + \Gamma^2} \text{Im}\{E_x E_y^*\}. \tag{18}$$

The transferred power is determined only by the non-conservative piece ($\bar{\sigma}_{EO}^{NH}$) of the EO conductivity. For simplicity, we dropped the subscript $\omega$, and from here on the dynamic electric field is simply denoted by $\mathbf{E}$. Clearly, the sign of $p_{dis,EO}$ can be either positive, indicating absorption, or negative, indicating gain. The structure of $\bar{\sigma}_{EO}^{NH}$ shows considerable similarity to the response of a passive dissipative gyrotropic medium. However, here we can have gain because the linear EO effect does not trigger a corresponding "partner" diagonal (symmetric) response. This contrasts with passive systems, where the symmetric component of the response invariably ensures a positive value for $p_{dis,EO}$. For circular or elliptical polarization, the sign of the term $\text{Im}\{E_x E_y^*\}$ is governed by the polarization handedness. Therefore, owing to the NH electrooptic contribution, power can be transferred either from the wave to the material (resulting in absorption, $p_{dis,EO} > 0$), or from the material to the wave (yielding optical gain, $p_{dis,EO} < 0$). It is relevant to note that $p_{dis,EO}$ is combined with an additional term $p_{dis,D} > 0$ due to the linear Drude response. This second term always generates dissipation.

## B. Full optical response

The full electromagnetic response of the material can be characterized by a permittivity tensor $\bar{\varepsilon}(\omega) = \bar{\varepsilon}_b + i\bar{\sigma}/\omega$, with $\bar{\sigma}(\omega) = \bar{\sigma}_D(\omega) + \bar{\sigma}_{EO}^H + \bar{\sigma}_{EO}^{NH}(\omega)$ describing the response of free electrons, and $\bar{\varepsilon}_b$ describing the permittivity response of bound electrons. Here $\bar{\sigma}_D(\omega)$ is the usual linear Drude response:

$$\bar{\sigma}_D(\omega) = \frac{\varepsilon_0 \omega_p^2}{\Gamma - i\omega} \bar{1} . \tag{19}$$

For simplicity, we ignore here the typical anisotropic response of materials with $\mathcal{D}_3$ symmetry, and we take $\bar{\varepsilon}_b$ as a diagonal matrix of the form $\bar{\varepsilon}_b = \varepsilon_0 \varepsilon_{diel} \bar{1}$. Furthermore, it is assumed that the response of bound electrons is little affected by the static electric bias. Furthermore, we also neglect the chiral response inherent to materials with $\mathcal{D}_3$ symmetry, which is associated with natural optical activity. Under these conditions, the permittivity takes the form $\bar{\varepsilon}(\omega) = \varepsilon_0 \left( \varepsilon_{diel} \bar{1} + \bar{\chi}_D(\omega) + \bar{\chi}_{EO}^H(\omega) + \bar{\chi}_{EO}^{NH}(\omega) \right)$, where the electric susceptibilities $\bar{\chi}_\alpha(\omega) = i\bar{\sigma}_\alpha/(\varepsilon_0 \omega)$ model the different components of the response of free electrons response ($\alpha = $ D, EO-H, EO-NH). Doing this, the relative permittivity tensor of the material can be written as:

$$\frac{\bar{\varepsilon}(\omega)}{\varepsilon_0} = \varepsilon_{diag} \bar{1} + i\varepsilon_g \hat{\mathbf{z}} \times \bar{1}, \qquad \text{with} \tag{20a}$$

$$\varepsilon_{\text{diag}}(\omega) = \varepsilon_{\text{diel}} - \frac{\omega_p^2}{\omega^2 + i\Gamma\omega} \quad \text{and} \quad \varepsilon_g(\omega) = \frac{\omega_0 \Gamma}{\omega}\left(\frac{2}{\Gamma} + \frac{i\omega + \Gamma}{\omega^2 + \Gamma^2}\right). \quad (20b)$$

For the sake of simplicity, we will omit the dispersion of $\varepsilon_{\text{diel}}$, but we will incorporate dissipation effects due to phonon coupling, represented by a constant imaginary component: $\varepsilon_{\text{diel}} = \varepsilon'_{\text{diel}} + i\varepsilon''_{\text{diel}}$.

Next, we characterize the plane wave modes in the material for propagation along the trigonal axis. The dispersion equation for this case can be simplified to $k_z^2 - (\omega/c)^2(\varepsilon_{\text{diag}} \pm \varepsilon_g) = 0$. Solving this dispersion equation yields the following solutions:

$$\text{Eigenmode 1:} \quad k_{z,1} = \frac{\omega}{c}\sqrt{\varepsilon_{\text{eff},1}} \quad \text{with} \quad \varepsilon_{\text{eff},1} = \varepsilon_{\text{diag}} + \varepsilon_g, \quad \mathbf{E}_1 \sim \hat{\mathbf{x}} + i\hat{\mathbf{y}}$$
$$\text{Eigenmode 2:} \quad k_{z,2} = \frac{\omega}{c}\sqrt{\varepsilon_{\text{eff},2}} \quad \text{with} \quad \varepsilon_{\text{eff},2} = \varepsilon_{\text{diag}} - \varepsilon_g, \quad \mathbf{E}_2 \sim \hat{\mathbf{x}} - i\hat{\mathbf{y}} \quad (21)$$

Therefore, the bulk material supports two eigenmodes: eigenmode 1 with positive handedness with respect to the +z-axis and eigenmode 2 with negative handedness with respect to +z.

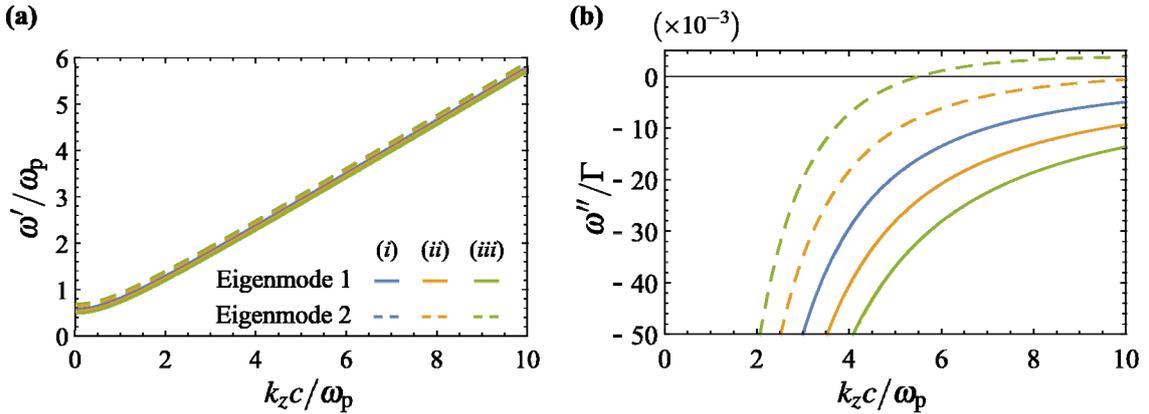

**Fig. 2**. Photonic band diagram of a conductive material with $\mathcal{D}_3$ symmetry biased along the trigonal axis (z-direction). The propagation is along z. (a) Real and (b) imaginary parts of the oscillation frequencies of the two eigenmodes as a function of $k_z$ for $\varepsilon_{\text{diel}} = 3$, $\omega_p/(2\pi) = 5$ THz, $\Gamma = 0.03\omega_p$, and (i) $\omega_0/\omega_p = 0$, (ii) $\omega_0/\omega_p = 0.15$, (iii) $\omega_0/\omega_p = 0.3$. For $\omega_0/\omega_p = 0$, the solid and dashed blue curves are coincident in both panels.

Figure 2 shows the photonic band diagrams ($\omega = \omega' + i\omega''$ vs $k \equiv k_z$) for different values of $\omega_0$. In the absence of bias ($\omega_0 = 0$), the material supports the propagation of two degenerate modes [see blue lines in Fig. 2(a)-(b)]. Differently, under a static electric bias ($\omega_0 \neq 0$), two circularly polarized eigenmodes with opposite handedness propagate in the material [see orange and green solid and dashed lines in Fig. 2(a)-(b)]. Notably, for the eigenmode 2 (dashed lines) an increase of $\omega_0$ diminishes $|\omega''|$ –corresponding to a larger lifetime of the natural mode– and may even lead to oscillations that grow exponentially with time ($\omega'' > 0$) [see Fig. 2(b)]. These exponentially growing oscillations (wave instabilities) arise from the non-Hermitian electrooptic effect, which becomes significant for large enough values of $|\omega_0|$. Quite differently, for eigenmode 1 (solid lines) $|\omega''|$ increases with $\omega_0$, so that the lifetime becomes shorter.

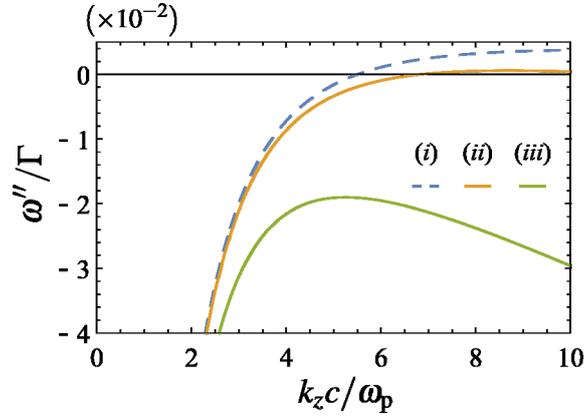

**Fig. 3**. Imaginary parts of the oscillation frequency of the eigenmode 2 as a function of $k_z$ for $\varepsilon'_{\text{diel}} = 3$, $\omega_0/\omega_p = 0.3$, and $\Gamma = 0.03\omega_p$. (i) $\varepsilon''_{\text{diel}} = 0$; (ii) $\varepsilon''_{\text{diel}} = 1\times10^{-4}$; (iii) $\varepsilon''_{\text{diel}} = 1\times10^{-3}$.

As expected, accounting for losses in the dielectric response ($\varepsilon''_{\text{diel}} \neq 0$) diminishes or even suppress the gain [see Fig. 3]. Finally, it should be noted that by reversing the sign of $\omega_0$, which according to Eq. (16) implies a reversal of the sign of either $E_0$ or $D$, the roles of the two eigenmode polarizations are interchanged.

*C. Stability*

Let us analyze the conditions to have a stable material response. The material is stable when the bulk modes with an arbitrary real-valued $k_z$, are associated with eigenfrequencies in the lower-half frequency plane ($\omega'' < 0$). To study this point, first we find the frequency $\omega$ where the dispersion diagram ($\omega = \omega(k_z)$) crosses the real-frequency axis. As $k_z = (\omega/c)\sqrt{\varepsilon_{diag} \pm \varepsilon_g}$, it follows that the crossing occurs for a frequency $\omega$ such that $\text{Im}\{\varepsilon_{diag} \pm \varepsilon_g\} = 0$. Solving this equation, it is found that $\omega$ is linked to the gain term ($\omega_0$) as:

$$|\omega_0| = \frac{1}{\omega \Gamma}\left(\varepsilon''_{diel}\omega(\omega^2 + \Gamma^2) + \Gamma \omega_p^2\right). \tag{22}$$

In order that $\omega(k_z)$ crosses the real-frequency axis exactly at the point $\omega$ the gain must be as in the above formula. The minimum gain to have a crossing can be found by minimizing the above expression with respect to $\omega$. The minimum occurs at the frequency $\omega_{th} = \left(\dfrac{\Gamma \omega_p^2}{2\varepsilon''_{diel}}\right)^{1/3}$. Thus, the threshold gain is given by:

$$|\omega_{0,th}| = \frac{1}{\omega_{th} \Gamma}\left(\varepsilon''_{diel}\omega_{th}(\omega_{th}^2 + \Gamma^2) + \Gamma \omega_p^2\right). \tag{23}$$

For $|\omega_0| < |\omega_{0,th}|$ the material response is unconditionally stable. When the gain matches the threshold value, the band diagram reaches the real-frequency axis exactly at the point $\omega' = \omega_{th}$. Note that for $\varepsilon''_{diel} = 0$, $\omega_{th} \to \infty$ and $|\omega_{0,th}| = 0$, indicating no instability threshold. This means that for $\varepsilon''_{diel} = 0$ and sufficiently large frequencies, the system can always become unstable, consistent with the results shown in Fig. 2(b).

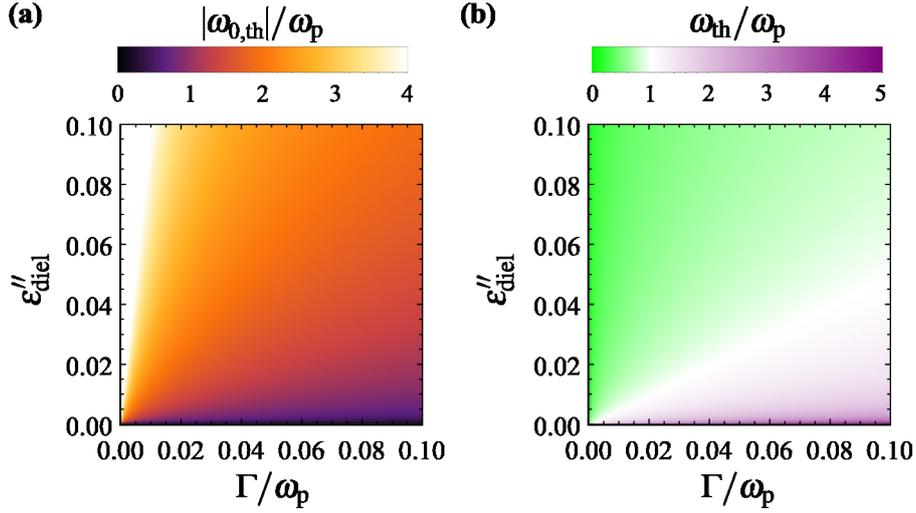

**Fig. 4.** (a) $|\omega_{0,\text{th}}|$ and (b) $\omega_{\text{th}}$ as a function of the scattering rate $\Gamma$ and of the imaginary part of the dielectric response $\varepsilon''_{\text{diel}}$.

Figure 4(a) shows the dependence of $|\omega_{0,\text{th}}|$ on the scattering rate $\Gamma$ and on the imaginary part of the dielectric response $\varepsilon''_{\text{diel}}$. It is evident from Fig. 4(a) that reducing $\varepsilon''_{\text{diel}}$ results in a decrease of $|\omega_{0,\text{th}}|$, whereas decreasing $\Gamma$ leads to an increase of $|\omega_{0,\text{th}}|$. Note that $|\omega_{0,\text{th}}| \sim 1/\Gamma$, and hence, for $\Gamma = 0$, $|\omega_{0,\text{th}}| \to \infty$.

Figure 4(b) shows how the threshold oscillation frequency $\omega_{\text{th}}$ (i.e., the instability frequency when the gain parameter matches the threshold) varies with $\Gamma$ and $\varepsilon''_{\text{diel}}$. For $\varepsilon''_{\text{diel}} < \Gamma/(2\omega_p)$ the instability frequency is larger than $\omega_p$ [purple region in Fig. 4(b)], whereas for $\varepsilon''_{\text{diel}} > \Gamma/(2\omega_p)$ it is smaller than $\omega_p$ [green region in Fig. 4(b)].

### D. Transmission and absorption spectra

To study the impact of the non-Hermitian EO effect on the optical response of the 3D material, next we consider that a slab of the material of thickness $d$ is illuminated by a circularly polarized electromagnetic wave. We assume that the material slab interfaces are perpendicular to the trigonal axis ($z$-direction). As before, the static electric bias ($\mathbf{E}_0$) is applied along $z$-direction, and the wave propagates along the direction of the bias (normal incidence). The material is surrounded by a vacuum.

For normal incidence, the gyrotropic material behaves as a standard dielectric with permittivity $\varepsilon_{\text{eff},1}$ ($\varepsilon_{\text{eff},2}$) for an incident electric field polarized as $\mathbf{E}_1 \sim \hat{\mathbf{x}} + i\hat{\mathbf{y}}$ ($\mathbf{E}_2 \sim \hat{\mathbf{x}} - i\hat{\mathbf{y}}$) [Eq. (21)]. Thus, the reflection and transmission coefficients take the form [52]:

$$R = \frac{i(\eta_0^2 - \eta^2)\sin(\beta d)}{2\eta_0\eta \cos(\beta d) - i(\eta_0^2 + \eta^2)\sin(\beta d)}, \tag{24a}$$

$$T = \frac{2\eta_0\eta}{2\eta_0\eta \cos(\beta d) - i(\eta_0^2 + \eta^2)\sin(\beta d)}, \tag{24b}$$

where $\eta = \eta_0 \sqrt{1/\varepsilon_{\text{eff}}}$ and $\beta = \frac{\omega}{c}\sqrt{\varepsilon_{\text{eff}}}$ are the intrinsic impedance and propagation constant of the equivalent medium ($\varepsilon_{\text{eff},1}$ or $\varepsilon_{\text{eff},2}$), and $\eta_0$ is the intrinsic impedance of the vacuum. The reflectance $\mathcal{R}$ and transmittance $\mathcal{T}$ can be written in terms of the reflection and transmission coefficients as $\mathcal{R} = |R|^2$ and $\mathcal{T} = |T|^2$. Finally, the absorptance $\mathcal{A}$ can be calculated from $\mathcal{A} = 1 - \mathcal{R} - \mathcal{T}$.

For an incoming wave that propagates along the +z direction, an incident field of the type $\mathbf{E}_1 \sim \hat{\mathbf{x}} + i\hat{\mathbf{y}}$ corresponds to a right-circularly polarized (RCP) wave, whereas an incident field of the type $\mathbf{E}_2 \sim \hat{\mathbf{x}} - i\hat{\mathbf{y}}$ corresponds to a left-circularly polarized (LCP) wave. Conversely, when the incoming wave propagates along the –z direction, illuminating the opposite face of the material slab, the mode $\mathbf{E}_1 \sim \hat{\mathbf{x}} + i\hat{\mathbf{y}}$ is associated with an LCP wave and the mode $\mathbf{E}_2 \sim \hat{\mathbf{x}} - i\hat{\mathbf{y}}$ is associated with a RCP wave. Thus, the transmittance for an RCP wave is $\mathcal{T} = \mathcal{T}_1$ when the direction of the incoming wave is +z, and $\mathcal{T} = \mathcal{T}_2$ when the direction of the incoming is –z. When the transmittances of modes 1 and 2 are different ($\mathcal{T}_1 \neq \mathcal{T}_2$) the system exhibits nonreciprocity [53].

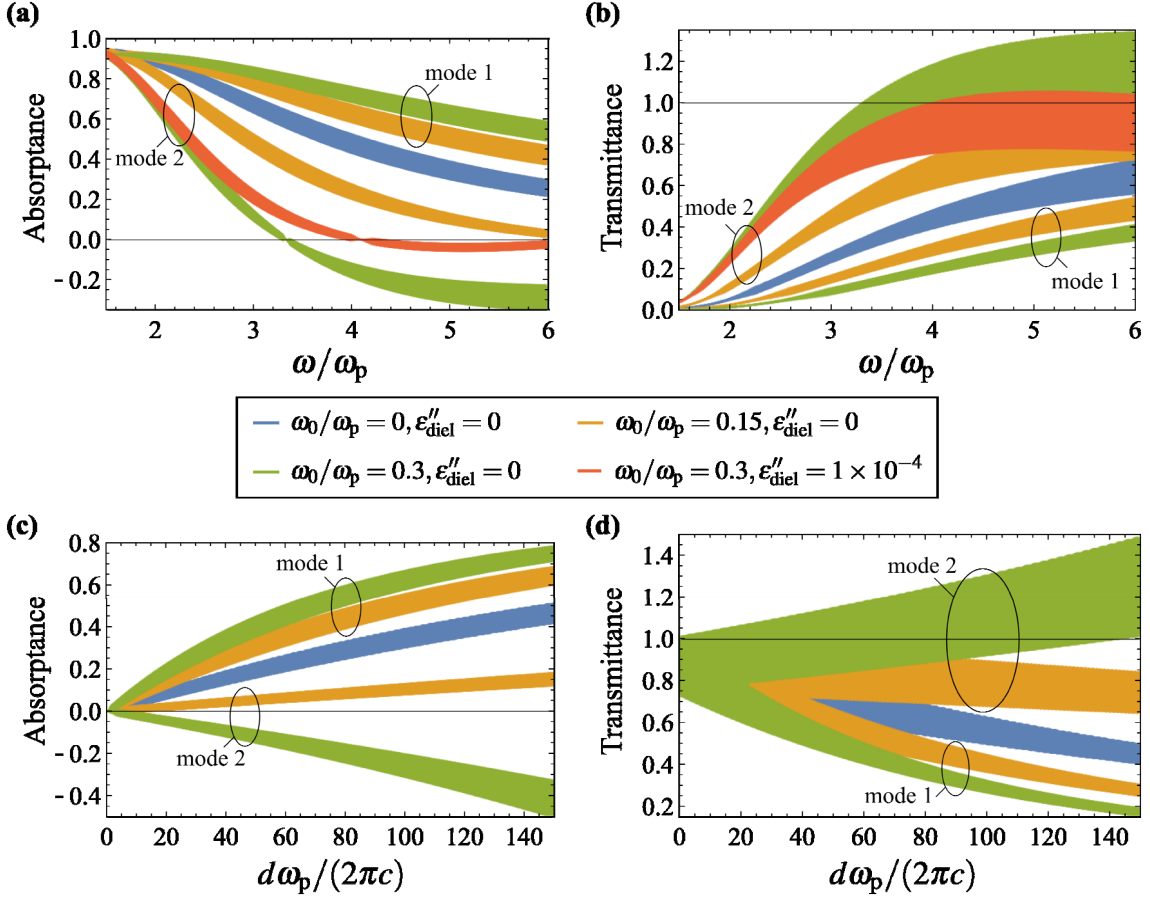

**Fig. 5**. (a) Absorptance and (b) transmittance as a function of the normalized frequency for an incident wave polarized as either mode 1 or mode 2. The normalized thickness of the material slab is $d\,\omega_\mathrm{p}/(2\pi c)=100$. (c) Absorptance and (d) transmittance as a function of the normalized thickness of the material slab for an incident wave polarized as either mode 1 or mode 2. The normalized frequency is $\omega/\omega_\mathrm{p}=5$. In all the panels, the material slab is characterized by $\varepsilon'_\mathrm{diel}=3$, $\omega_\mathrm{p}/(2\pi)=5\,\mathrm{THz}$, and $\Gamma=0.03\omega_\mathrm{p}$. For $\omega_0/\omega_\mathrm{p}=0$ (blue curves) the absorptance and transmittance are polarization independent.

In Figure 5, we depict the absorption and transmission characteristics of the material, presenting their frequency dependence in panels (a) and (b), and their sensitivity to the variation of the material thickness in panels (c) and (d). The curves across all panels of Fig. 5 exhibit prominent Fabry-Perot oscillations, where the absorption and transmission peaks of the unbiased system (blue curves) are determined by $(\omega/c)\sqrt{\varepsilon'_\mathrm{diel}}\,d=n\pi$. Due to the very close proximity between adjacent Fabry-Perot resonances within the plotted scales, the curves exhibit a multitude of tightly packed oscillations. This creates a visual perception of curves with large thickness dictated by the minima and maxima of the oscillations. The thickness of the curves is primarily influenced by the dielectric permittivity contrast between the material and the surrounding air region.

Figure 5(a) shows the absorptance of the electrically-biased 3D slab as a function of frequency, for different values of $\omega_0/\omega_p$ and $\varepsilon''_{\text{diel}}$, and for a normalized thickness $d\,\omega_p/(2\pi c) = 100$. When $\omega_0/\omega_p = 0$ (i.e., for an unbiased material or for a material without electrooptic response), the absorptances $\mathcal{A}_1$ and $\mathcal{A}_2$ of modes 1 and 2 are identical due to the rotational symmetry of the permittivity tensor [see Eq. (20)]. When $\omega_0/\omega_p \neq 0$, the material EO response combined with the intrinsic material loss induce circular dichroism, resulting in unequal absorptance for incident waves with opposite handedness ($\mathcal{A}_1 \neq \mathcal{A}_2$). Crucially, for sufficiently large values of $\omega_0/\omega_p$, the absorptance for incident waves polarized as $\mathbf{E}_2$ diminishes and may even become negative ($\mathcal{A}_2 < 0$) [see Fig. 5(a)], indicating the emergence of optical gain. In contrast, the absorptance for waves with the $\mathbf{E}_1$ polarization rises as $\omega_0/\omega_p$ increases. The polarization-dependent optical gain is rooted in the non-Hermitian EO effect, the strength of which becomes significant for large enough $\omega_0/\omega_p$.

Remarkably, for large values of $\omega_0/\omega_p$ the non-Hermitian EO response of the material may also lead to gain in transmission for waves with the $\mathbf{E}_2$-polarization. This is shown in Fig. 5(b), where the transmittance $\mathcal{T}_2$ can surpass unity for high enough $\omega_0/\omega_p$. Conversely, the transmittance for waves with the $\mathbf{E}_1$-polarization decreases as $\omega_0/\omega_p$ increases, and never exceeds unity.

As expected, an increase in the value of the dielectric losses ($\varepsilon''_{\text{diel}}$) leads to a decrease in the gain, both in terms of absorption and transmission, potentially leading to its suppression [see red curves in Fig. 5(a)-(b)].

Figures 5(c)-(d) depict the absorptance and transmittance of the electrically-biased 3D material as a function of the slab thickness $d$, for different values of $\omega_0/\omega_{\text{p}}$ and a fixed frequency of operation $\omega = 5\omega_{\text{p}}$. It is evident from Fig. 5(c) that increasing the slab thickness leads to a larger $|\mathcal{A}|$ (dissipation if $\mathcal{A} > 0$ or gain if $\mathcal{A} < 0$) for both polarization handednesses. Moreover, an increase in slab thickness results in higher (lower) transmittance values for mode 2 (mode 1) [see Fig. 5(d)].

## IV. NON-HERMITIAN LINEAR ELECTROOPTIC EFFECT IN TELLURIUM

### A. *Optical response with density-functional-theory (DFT) calculations*

Tellurium (Te) is a promising material for the observation of the non-Hermitian linear EO effect. Under typical conditions, tellurium is a slightly *p*-doped chiral semiconductor with a gyrotropic noncentrosymmetric crystal structure [40]. However, it is noteworthy that under certain conditions, *n*-doping can also be achieved [54-58]. Interestingly, tellurium showcases the ability to exhibit large Berry curvature dipoles (*D*) near the Weyl points [40]. Various phenomena observed in tellurium, such as kinetic Faraday rotation [40-43], current-induced magnetization (kinetic magnetoelectric effect) [40, 44-45], and the circular photogalvanic effect [40, 59-63], find their origin in the existence of the large Berry curvature dipoles.

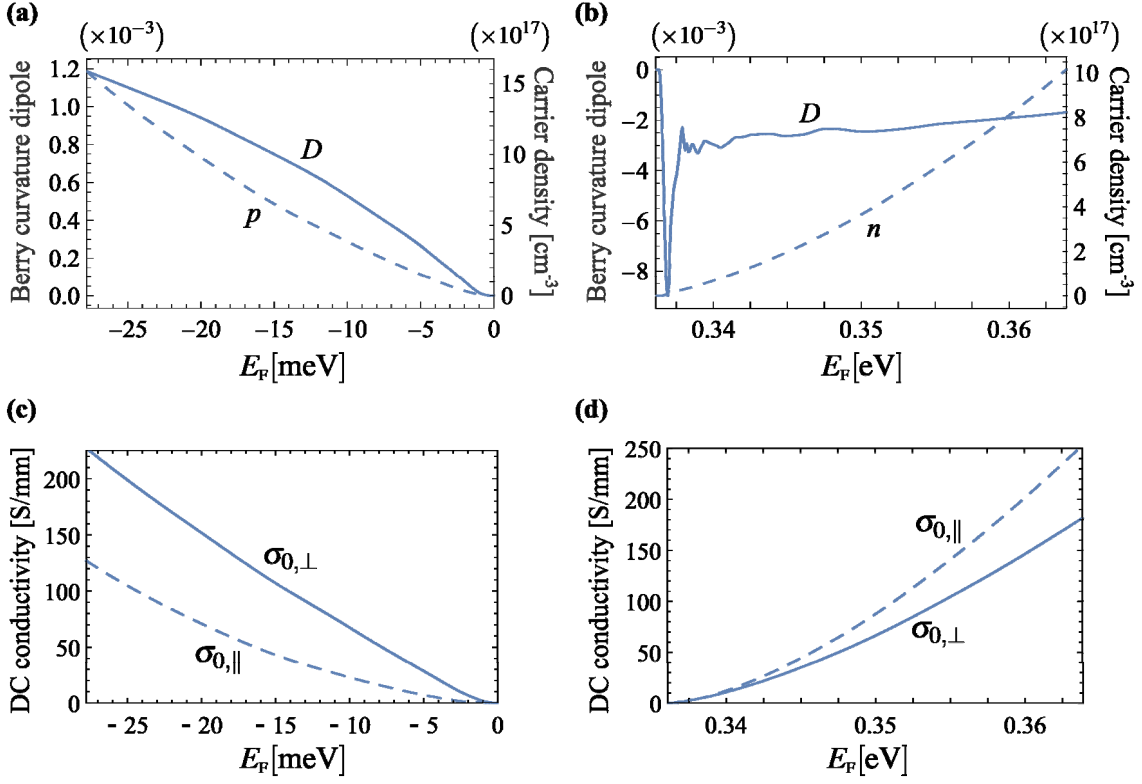

**Fig. 6**. (a) and (b) Berry curvature dipole *D* and carrier density (*p* or *n*) as a function of the Fermi energy $E_F$ (measured from the top of the upper valence band) for (a) *p*-doped Te and (b) *n*-doped Te. (c) and (d) DC conductivities as a function of the Fermi energy $E_F$ (assuming a scattering relaxation time $\tau = 0.64$ ps [42]) for (c) *p*-doped Te and (d) *n*-doped Te.

Figures 6(a)-(b) illustrate the relation between the Berry curvature dipole (*D*) and the carrier density (hole density *p* in (a), and electron density *n* in (b)) as a function of the Fermi energy $E_F$. In addition, the dependence of the DC conductivities on $E_F$ is shown in Figs. 5(c)-(d). The data is obtained from first-principles DFT calculations [40]. Notably, Fig. 6 reveals that *n*-doping of tellurium leads to larger magnitudes of the Berry curvature dipole compared to the more common *p*-doping scenario. Crucially, these large Berry curvature dipoles in *n*-doped tellurium emerge under conditions of low DC conductivities [see Fig. 6(d)]. Low conductivities are advantageous as they minimize Joule loss arising from the DC bias.

Elemental tellurium is a nonmagnetic semiconductor that forms two enantiomorphic trigonal structures with space groups P3$_1$21 (or $\mathcal{D}_3^4$; right-handed structure) and P3$_2$21 (or $\mathcal{D}_3^6$; left-handed structure), belonging to the 32-point group [40, 64] discussed in Sec. III. Besides the screw symmetry along the trigonal *z*-axis, the crystal structure also displays twofold rotational symmetry along three axes perpendicular to *z*. It is noteworthy that this screw symmetry along the trigonal *z*-axis is directly linked to the helical-type current orbits discussed in Sect. II.

The Berry curvature dipole tensor of tellurium is equivalent to that given in Eq. (14) [40]. However, due to the anisotropy of the material, its electromagnetic behavior cannot be described by the relative permittivity tensor (20). Instead, tellurium is characterized by the following relative permittivity tensor:

$$\frac{\bar{\varepsilon}(\omega)}{\varepsilon_0} = \varepsilon_\perp \bar{\mathbf{1}}_t + i\varepsilon_g \hat{\mathbf{z}} \times \bar{\mathbf{1}}_t + \varepsilon_\parallel \hat{\mathbf{z}} \otimes \hat{\mathbf{z}}, \tag{25}$$

with $\bar{\mathbf{1}}_t = \bar{\mathbf{1}} - \hat{\mathbf{z}} \otimes \hat{\mathbf{z}}$, $\varepsilon_\perp(\omega) = \varepsilon_{\text{diel},\perp} - \frac{\omega_{p,\perp}^2}{\omega^2 + i\Gamma\omega}$, $\varepsilon_\parallel(\omega) = \varepsilon_{\text{diel},\parallel} - \frac{\omega_{p,\parallel}^2}{\omega^2 + i\Gamma\omega}$ and $\varepsilon_{\text{diel},l} = \varepsilon'_{\text{diel},l} + i\varepsilon''_{\text{diel},l}$ (with $l = \perp, \parallel$ denoting the directions parallel and perpendicular to the trigonal axis, respectively). As in Sec. III, for a static electric bias along the trigonal *z*-axis, $\varepsilon_g(\omega) = \frac{\omega_0 \Gamma}{\omega}\left(\frac{2}{\Gamma} + \frac{i\omega + \Gamma}{\omega^2 + \Gamma^2}\right)$. For simplicity, in Eq. (25) we neglect the chiral response of tellurium which is responsible for its natural optical activity [42, 65]. In what follows, relying on Eq. (25), as well as on a first-principles calculations that provide the Berry curvature dipole *D* and the DC conductivity [40], we investigate several optical effects that arise from the gyrotropic and non-Hermitian linear EO responses of tellurium. For propagation along the *z* direction, the equivalent permittivity for modes 1 and 2 are now $\varepsilon_{\text{eff},1} = \varepsilon_\perp + \varepsilon_g$ and $\varepsilon_{\text{eff},2} = \varepsilon_\perp - \varepsilon_g$.

### B. Kinetic Faraday effect, dichroism and optical gain

To begin with, we investigate the rotation of light polarization in electrically-biased tellurium. This current-induced Faraday effect was experimentally observed for the first time in Ref. [41] at a temperature of 77K, and subsequently, new measurements were reported in Ref. [42]. We consider a system formed by a block of tellurium biased by a static electric field oriented along the *z*-direction (trigonal axis) in between two linear polarizers [see Fig. 7(a)]. The linear polarizers consist of wire grids designed to fully absorb the electric field component parallel to the wires and allow the full transmission of the orthogonal component. The wires orientation of the input WGP is fixed along the *x*-direction. As a result, for an incident wave propagating from the right to the left ($+z$-direction), the input WGP ensures that the field that illuminates the tellurium slab is oriented along the *y*-direction. On the other hand, the output WGP is free to rotate in the *xoy* plane. The angle $\alpha$ defines the orientation of the wires of the output WGP [see Fig. 7(a)]. The output WPG fully suppresses the electric field component parallel to the wire axis direction defined by $\hat{\mathbf{u}}_\alpha = \cos(\alpha)\hat{\mathbf{x}} + \sin(\alpha)\hat{\mathbf{y}}$, while allowing the orthogonal component to pass through unchanged. We assume an operational frequency $f = 28.3 \text{ THz}$ and a scattering relaxation time $\tau = 0.64 \text{ ps}$, consistent with the parameters used in Ref. [42]. Additionally, the thickness of the Te block is $d = 12.5 \text{ mm}$, as in Ref. [42]. Moreover, we estimate that for $f = 28.3 \text{ THz}$ the dielectric response $\varepsilon_{\text{diel},\perp} = 21.9 + 5.8 \times 10^{-4} i$. The dielectric response was determined through extrapolation from the experimental data presented in Ref. [66], using a fit with a Lorentz dispersion equation.

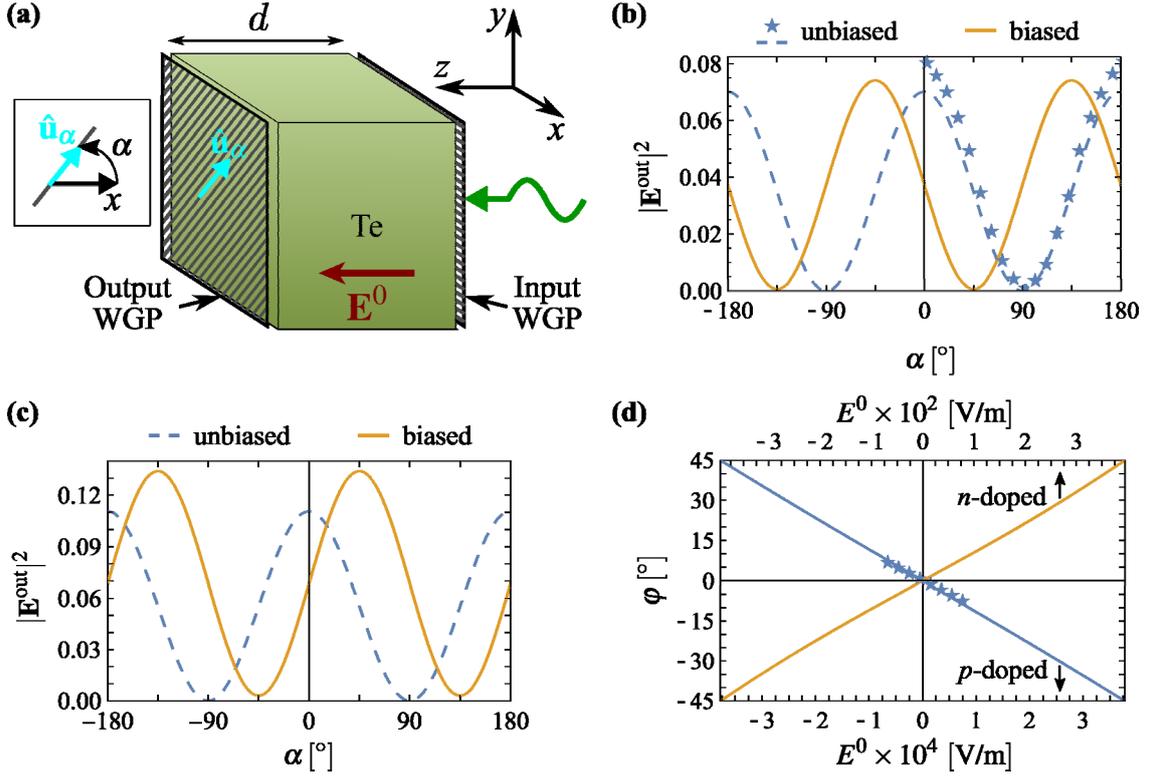

**Fig. 7**. (a) Electrically-biased block of Te placed in between two linear polarizers. (b) and (c) Light intensity ($\left|\mathbf{E}^{\mathrm{out}}\right|^2$) at the output (i.e., after the polarizer) as a function of orientation of the output WGP. (b) For $p$-doped Te with $p = 4 \times 10^{16}$ cm$^{-3}$, $D \approx 8.886 \times 10^{-5}$, and $\omega_{\mathrm{p},\perp}/(2\pi) \approx 5.948$ THz, without an static electric bias (blue dashed line and star symbols), and with a bias $E_0 = 37.793$ V/mm. (c) For $n$-doped Te with $n = 0.737 \times 10^{16}$ cm$^{-3}$, $D \approx -8.99 \times 10^{-3}$, and $\omega_{\mathrm{p},\perp}/(2\pi) \approx 1.857$ THz, without an static electric bias (blue dashed line), and with a bias $E_0 = 0.3945$ V/mm. (d) Electric field rotation ($\varphi = \alpha_1 - \alpha_0$) as a function of the static electric bias $E_0$. $\alpha_1$ ($\alpha_0$) corresponds to the angle $\alpha$ that provides maximum light intensity after the output WGP for a system biased with a static electric field $E_0$ (for an unbiased system); Blue solid line and star symbols: $p$-doped Te slab with the same parameters as in panel (b); Orange solid line: $n$-doped Te slab with the same parameters as in panel (c). In panels (b), (c) and (d) the frequency of operation is $f = 28.3$ THz, the thickness of the Te block is $d = 12.5$ mm, $\varepsilon_{\mathrm{diel},\perp} = 21.9 + 5.8 \times 10^{-4} i$, and $\Gamma = 1.5625 \times 10^{12}$ rad/s. The blue star symbols in panels (b) and (d) correspond to the experimental results of Ref. [42].

Figure 7(b) depicts the light intensity at the output (after the polarizer) as a function of the angle $\alpha$ for (*i*) an unbiased $p$-doped Te block and (*ii*) a biased $p$-doped Te block with $E_0 = 37.793$ V/mm (which corresponds to $\left|\omega_0/\omega_{\mathrm{p},\perp}\right| \approx 2.404 \times 10^{-3}$). The transmission and reflection coefficients for an incoming wave with linear polarization can be found from Eqs. (24), by writing the incident wave as a superposition of two circularly polarized waves with opposite handedness.

Notably, our results (blue dashed line) and the experimental results of Ref. [42] (blue star symbols) for the unbiased Te slab exhibit a qualitatively good agreement. Note that the experimental results of [42] already exclude the field polarization rotation arising from the chiral response of tellurium. The $E_0$ values for these experimental results are estimated from the data available in Ref. [42] using $E_0 = j_z/(p\mu e)$, with $j_z$ the electric current density along the z-direction and $\mu$ the hole mobility. Interestingly, in this example the output light intensity $|\mathbf{E}^{out}|^2$ for the biased system exhibits a slight increase compared to the unbiased case. Owing to the presence of Fabry-Pérot resonances, the output intensity $|\mathbf{E}^{out}|^2$ exhibits strong sensitivity to both the slab thickness $d$ and the magnitude of the static electric field bias $E_0$. Even slight adjustments of $d$ and/or $E_0$ can lead to a substantial change in the output amplitude. The system could be easily tuned to have an $|\mathbf{E}^{out}|^2$ for the biased system larger (or smaller) than that for the unbiased system.

More interestingly, Fig. 7(b) shows that for $E_0 = 37.793$ V/mm the polarization of an electromagnetic wave passing through the biased *p*-doped Te slab is rotated by $\pm 45°$. This suggests the potential for creating an optical isolator using an electrically-biased (along the trigonal z-axis) slab of *p*-doped Te. However, it is crucial to acknowledge that the required static electric bias magnitude is high (more than five times larger than the maximum $E_0$ employed in the experimental work [42]), probably rendering practical implementation unfeasible. One approach to circumvent this problem is to use materials with much larger Berry curvature dipoles [67]. Another possibility involves using a block of *n*-doped tellurium [54-58] instead of *p*-doped tellurium. Remarkably, *n*-doped tellurium can offer large values of the Berry curvature dipole $D$ with low DC conductivity [see Fig. 6]. Specifically, for *p*-doped tellurium and $p = 4 \times 10^{16}$ cm$^{-3}$, $D \approx 8.886 \times 10^{-5}$, $\sigma_{0,\perp} \approx 7.9$ S/mm, and $\sigma_{0,\parallel} \approx 1.178$ S/mm, whereas for *n*-doped tellurium and $n = 0.737 \times 10^{16}$ cm$^{-3}$, $D \approx -8.99 \times 10^{-3}$, $\sigma_{0,\perp} \approx 0.770$ S/mm, and $\sigma_{0,\parallel} \approx 1.229$ S/mm. However, it is important to note that for *n*-doped tellurium, when the optical frequency is non-negligible compared to the dominant interband transitions, the corrections to the semiclassical approximation used here can reduce the magnitude of $D$ [40].

In Fig. 7(c), we show how the output light intensity $|\mathbf{E}^{out}|^2$ varies with the angle $\alpha$ for (*i*) an unbiased *n*-doped Te block and (*ii*) a biased *n*-doped Te block with $E_0 = 0.3945$ V/mm (which corresponds to $|\omega_0/\omega_{p,\perp}| \approx 0.813 \times 10^{-2}$). By comparing Figs. 7(b) and (c), it becomes evident that the relationship between the sign of the static field bias $E_0$ and the direction of the polarization rotation in a biased *p*-doped tellurium slab is opposite to that observed in a biased *n*-doped tellurium slab. This is a consequence of the differently signed Berry curvature dipoles: positive for *p*-doped tellurium and negative for *n*-doped tellurium [see Fig. 6]. More importantly, the results of Fig. 7(c) demonstrate that a $45°$ polarization rotation in the *n*-doped Te slab can be achieved by applying a much lower electric bias (specifically, $E_0 = 0.3945$ V/mm), almost two orders of magnitude smaller than that needed for the same polarization rotation in the *p*-doped Te slab. Quite interestingly, $E_0 = 0.3945$ V/mm is almost eighteen times smaller than the maximum value of $E_0 = 7$ V/mm employed in the experiments reported in Ref. [42]. Thus, *n*-doped tellurium emerges as an interesting candidate for the practical realization of electrically-biased optical isolators.

The sharp difference in the required static electric bias ($E_0$) for achieving equivalent polarization rotation in *n*-doped and *p*-doped Te becomes even more evident in Fig. 7(d). The figure depicts the dependence of the angle of rotation $\varphi$ of the plane of polarization of $\mathbf{E}^{out}$ on the static electric bias $E_0$ for both *p*-doped and *n*-doped Te slabs. Notably, our theoretical results (blue solid line) and the experimental data from Ref. [42] (blue star symbols) for the *p*-doped Te slab concur very closely. Furthermore, as seen in Fig. 7(d), *p*-doped and *n*-doped Te slabs exhibit polarization rotations in opposite directions, in agreement with the results of Figs. 7(b) and (c). Finally, the results of Fig. 7(d) show that a *n*-doped Te slab requires electric bias magnitudes ($|E_0|$) almost one hundred times smaller than those needed by the *p*-doped Te to provide the same rotation of polarization $\varphi$. For example, for a 45° polarization rotation *p*-doped Te requires a bias of $|E_0| = 37.793$ V/mm, whereas *n*-doped Te only requires $|E_0| = 0.3945$ V/mm. The bias $|E_0| = 0.3945$ V/mm is only 5.64% of the maximum $E_0$ utilized in the experimental measurements of Ref. [42]. Thus, *n*-doped tellurium is a prime candidate for realizing electrically-biased optical isolators in practical scenarios.

Next, we investigate optical dichroism in an electrically-biased tellurium slab. As before, we suppose that the tellurium slab is subjected to a static electric bias ($\mathbf{E}_0$) along the *z*-direction (trigonal axis) and illuminated by circularly polarized electromagnetic waves propagating in the same *z*-direction.

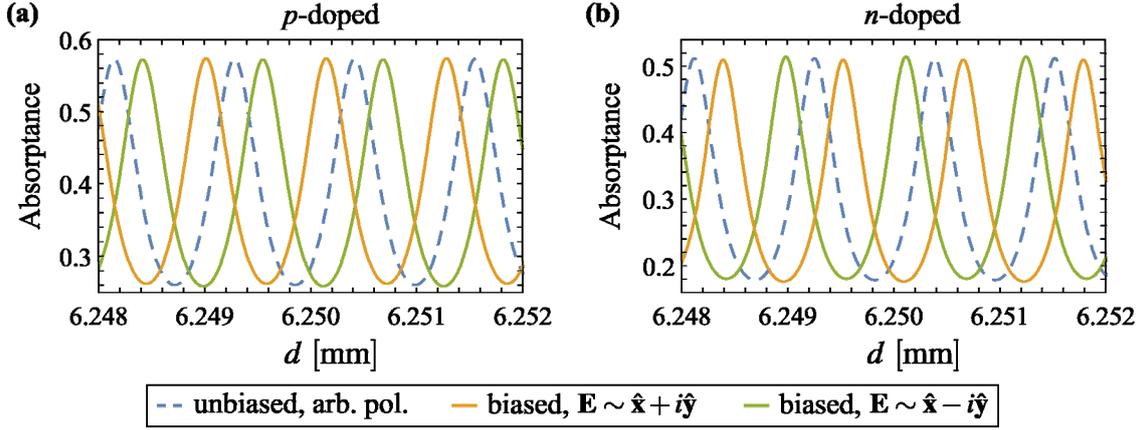

**Fig. 8**. Absorptance as a function of the thickness of the Te slab for a (a) *p*-doped and (b) *n*-doped material. Dashed blue lines: unbiased ($E_0 = 0$) Te for arbitrarily polarized incident waves; solid orange lines: biased Te for an incident wave polarized as mode 1 ($\mathbf{E}_1 \sim \hat{\mathbf{x}} + i\hat{\mathbf{y}}$); solid green lines: biased Te for an incident wave polarized as mode 2 ($\mathbf{E}_1 \sim \hat{\mathbf{x}} - i\hat{\mathbf{y}}$). (a) $p = 4 \times 10^{16}$ cm$^{-3}$ and $E_0 = 70$ V/mm (which corresponds to $|\omega_0/\omega_{p,\perp}| \approx 4.452 \times 10^{-3}$); (b) $n = 0.737 \times 10^{16}$ cm$^{-3}$ and $E_0 = 0.7$ V/mm (which corresponds to $|\omega_0/\omega_{p,\perp}| \approx 1.443 \times 10^{-2}$). The frequency of operation is $f = 28.3$ THz, $\varepsilon_{\text{diel},\perp} = 21.9 + 5.8 \times 10^{-4} i$, and $\Gamma = 1.5625 \times 10^{12}$ rad/s.

In Figure 8, we present the absorptance characteristics of unbiased (dashed curves) and electrically-biased (solid curves) *p*-doped (Fig. 8(a)) and *n*-doped (Fig. 8(b)) tellurium slabs as a function of slab thickness $d$ and for a fixed frequency $f = 28.3$ THz. It is evident from Fig. 8(a)-(b) that the absorptance curves display a Fabry-Perot oscillation pattern. The absorption peaks of the unbiased system (dashed curves) are determined by $(\omega/c)\sqrt{\varepsilon_{\text{diel},\perp}}\, d = n\pi$, where *n* is an integer.

For an unbiased system ($E_0 = 0$) and propagation along *z*-direction (trigonal axis), the permittivity of tellurium is isotropic, leading to polarization-independent absorptance [see blue dashed curves in Fig. 8(a)-(b); as before, we ignore chiral effects]. Notably, under a static electric bias ($E_0 \neq 0$), the absorptance becomes polarization-dependent, and the absorptance peaks undergo a relative shift with respect to the unbiased case. The absorptance peaks associated with $\mathbf{E}_1 \sim \hat{\mathbf{x}} + i\hat{\mathbf{y}}$ and $\mathbf{E}_2 \sim \hat{\mathbf{x}} - i\hat{\mathbf{y}}$ polarized incident waves undergo opposite shifts, with one polarization shifted rightward and the other shifted leftward. Therefore, electrically-biased tellurium exhibits optical dichroism. This optical dichroism arises from the combination of the gyrotropic Hermitian EO response with the intrinsic loss of Te and can become significant when $E_0$ (or $\omega_0/\omega_{p,\perp}$) reaches sufficiently high magnitudes. Within the parameter range considered in Fig. 8(a)-(b), the absorptances for modes 1 ($\mathcal{A}_1$) and 2 ($\mathcal{A}_2$) can differ by more than 55%. The results of Fig. 8(a)-(b) demonstrate that to achieve a comparable level of optical dichroism, the required magnitude of $E_0$ in the *n*-doped tellurium is two orders of magnitude lower than that needed in the *p*-doped tellurium. Notably, the magnitude of $E_0$ considered for the *n*-doped tellurium ($E_0 = 0.7$ V/mm) is one order of magnitude below the maximum $E_0$ used in the experiments of Ref. [42] (specifically, $E_0 = 7$ V/mm). Furthermore, it is important to highlight that the value of $E_0$ can be further reduced by increasing the thickness of the tellurium block.

To conclude, we demonstrate that due to the non-Hermitian linear EO effect, tellurium may also provide optical gain. Figures 9(a)-(b) display how absorptance varies with the magnitude of the static electric bias ($E_0$) for both *p*-doped (Fig. 9(a)) and *n*-doped (Fig. 9(b)) tellurium. Quite remarkably, Figs. 9(a)-(b) show that when $E_0$ surpasses a certain threshold, the absorptance for circularly polarized incident waves of a certain handedness can become negative [see green curve in Fig. 9(a) and the orange curve in Fig. 9(b)], indicating the emergence of optical gain. This gain regime emerges due to the non-conservative (transistor-type) light-matter interactions associated with the Berry curvature dipoles, which creates the opportunity to have optical gain controlled by the polarization handedness.

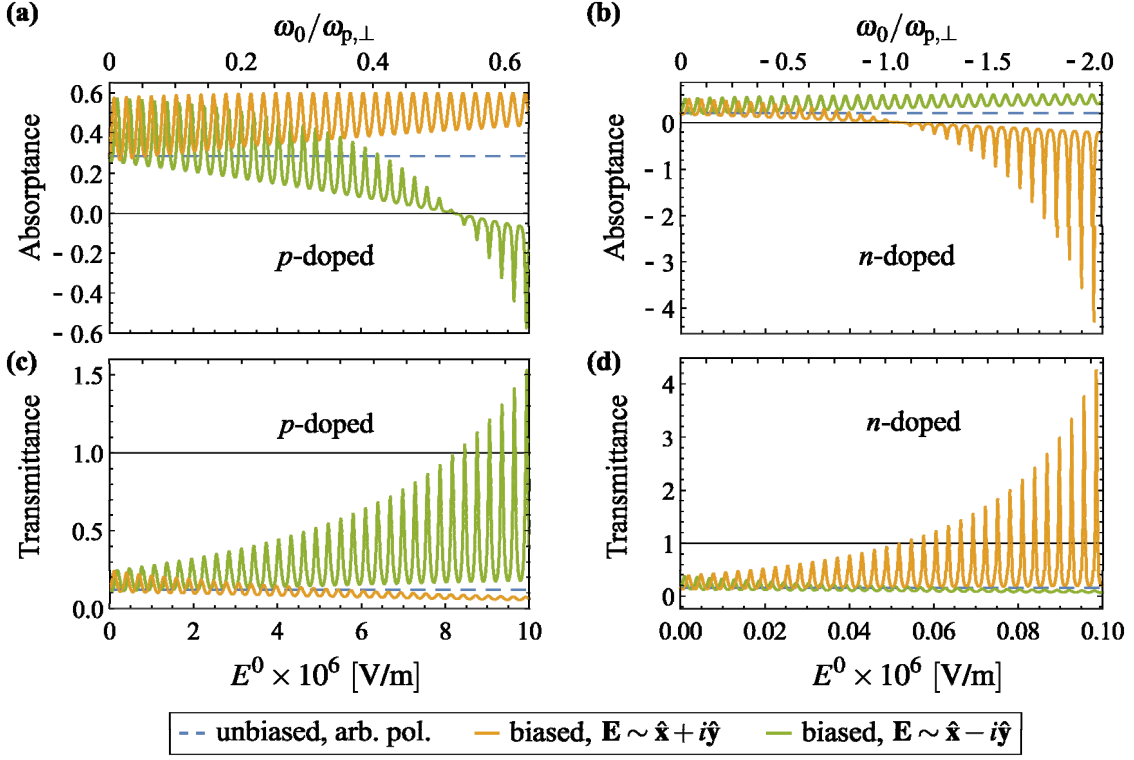

**Fig. 9**. (a)-(b) Absorptance and (c)-(d) transmittance as a function of the static electric field bias $E_0$ and $\omega_0/\omega_{p,\perp}$. (a) and (c) $p$-doped Te with $p = 4\times10^{16}$ cm$^{-3}$, (b) and (d) $n$-doped Te with $n = 0.737\times10^{16}$ cm$^{-3}$. Dashed blue lines: unbiased ($E_0 = 0$) Te for arbitrarily polarized incident waves; solid orange lines: biased Te for an incident wave polarized as mode 1 ($\mathbf{E}_1 \sim \hat{\mathbf{x}}+i\hat{\mathbf{y}}$); solid green lines: biased Te for an incident wave polarized as mode 2 ($\mathbf{E}_1 \sim \hat{\mathbf{x}}-i\hat{\mathbf{y}}$). The frequency of operation is $f = 28.3$ THz, the thickness of the Te slab is $d = 6.25$ mm, $\varepsilon_{\text{diel},\perp} = 21.9 + 5.8\times10^{-4}i$, and $\Gamma = 1.5625\times10^{12}$ rad/s.

For $p$-doped tellurium, $E_0 > 0$, and propagation along the $+z$-direction, the gain regime, associated with negative absorption, occurs for LCP incident waves [see green curve in Fig. 9(a)]. However, achieving this gain response in $p$-doped tellurium requires exceptionally high values of $E_0$, exceeding $8.24$ V/$\mu$m. In contrast, for $n$-doped tellurium the optical gain emerges for RCP incident waves [see orange curve in Fig. 9(b)]. Crucially, in the case of $n$-doped tellurium, the gain response is unlocked for much lower static electric bias $E_0$, specifically $E_0 > 52$ V/mm. It is worth noting that these $E_0$-thresholds are consistent with the values obtained using Eqs. (16) and (22). Therefore, the magnitude of $E_0$ required for achieving optical gain in $n$-doped tellurium is more than two orders of magnitude smaller than that needed in *p*-doped tellurium. Unfortunately, applying an $E_0$ of the order of 52 V/mm poses significant practical challenges. Such a value is seven times larger than the maximum value of $E_0$ employed in the experiments of Ref. [42]. In future work, will study other more practical configurations, or more practical materials with larger Berry curvature dipoles [67].

In Figs. 9(c)-(d), we depict the transmittance as a function of the static electric bias magnitude ($E_0$) for both *p*-doped (Fig. 9(c)) and *n*-doped (Fig. 9(d)) tellurium. Notably, these results show that for large enough $E_0$, the *p*-doped (*n*-doped) tellurium may also provide transmission gain as the transmittance for LCP (RCP) incident waves may exceed unity. Analogous to the absorptance case, *n*-doped tellurium requires an $E_0$ roughly a hundred times smaller to achieve the same transmission gain as with *p*-doped tellurium.

## V. CONCLUSIONS

In this work, relying on first principles density functional theory (DFT) and Boltzmann transport theory, we have conducted an in-depth theoretical analysis of the linear EO effect within low-symmetry three-dimensional (3D) conductive materials characterized by large Berry curvature dipoles. In line with Ref. [36], our study reveals the presence of two distinct contributions to the linear EO effect: one stemming from a gyrotropic Hermitian contribution linked to a conservative response, and another arising from a non-Hermitian electrooptic response characterized by its non-conservative nature. The gyrotropic Hermitian EO response may be pictured as being the result of helical-type microscopic currents induced by the static electric bias, leading to the generation of an internal static magnetic field. Remarkably, the Berry curvature dipole tensor may be understood as a magneto-electric coupling tensor that relates the equivalent internal magnetic field to the static electric bias.

Our analysis was centered on 3D materials with $\mathcal{D}_3$ symmetry when subjected to a static electric bias along the trigonal axis. It was demonstrated that such electrically biased materials may potentially provide negative absorption (i.e., optical gain) and transmission amplification. Finally, we have identified trigonal tellurium as a promising material for exploring and leveraging the non-Hermitian EO effect. It was demonstrated that the gyrotropic Hermitian response of tellurium may enable realizing electrically-biased electromagnetic isolators, as well as induce significant optical dichroism. Moreover, under extreme conditions of operation, the non-Hermitian electrooptic response of tellurium can lead to optical gain. In such scenarios, *n*-doped tellurium is especially noteworthy due to its larger Berry curvature dipoles and lower DC conductivity, as compared to the commonly used *p*-doped tellurium.

**Acknowledgments:** This work was partially funded by the IET under the A F Harvey Prize, by the Simons Foundation under the award 733700 (Simons Collaboration in Mathematics and Physics, "Harnessing Universal Symmetry Concepts for Extreme Wave Phenomena"), and by Instituto de Telecomunicações (IT) under project UIDB/50008/2020. T. A. M. acknowledges FCT for research financial support with reference CEECIND/04530/2017/CP1393/CT0004 (DOI 10.54499/CEECIND/04530/2017/CP1393/CT0004) under the CEEC Individual 2017, and IT-Coimbra for the contract as an assistant researcher with reference CT/No. 004/2019-F00069. T. G. R. acknowledges funding from FCT-Portugal through Grant No. CEECIND/07471/2022. S. L. acknowledges FCT and IT-Coimbra for the research financial support with reference DL 57/2016/CP1353/CT000. S. S. T. and I. S. acknowledge funding by MCIN/AEI/10.13039/501100011033 through Grant No. PID2021-129035NB-I00.